\documentclass[conference]{IEEEtran}
\IEEEoverridecommandlockouts

\usepackage{amsmath,amssymb,amsfonts}
\usepackage{cite}
\usepackage{graphicx}
\usepackage{booktabs}
\usepackage[caption=false,font=footnotesize]{subfig}
\usepackage{url}
\usepackage{xcolor}
\usepackage{algorithm}
\usepackage{algpseudocode}
\algrenewcommand\algorithmicrequire{\textbf{Input:}}
\algrenewcommand\algorithmicensure{\textbf{Output:}}
\usepackage{enumitem}
\usepackage{textcomp}
\usepackage[hidelinks]{hyperref}

\title{A Closed-Loop Thermal Dynamic Model for AI Data Center Cooling Load Simulation
\thanks{This work was supported by the Power Systems Engineering Research Center (PSERC) under project T-77. Source code: \url{https://github.com/cletuzz00/dc-cooling-thermal-model}}}

\author{
\IEEEauthorblockN{Cletus Ngwerume}
\IEEEauthorblockA{\textit{Michigan Technological University}\\
Houghton, MI, USA\\
ctngweru@mtu.edu}
\and
\IEEEauthorblockN{Lang Tong}
\IEEEauthorblockA{\textit{Cornell University}\\
Ithaca, NY, USA\\
lt35@cornell.edu}
\and
\IEEEauthorblockN{Chee-Wooi Ten}
\IEEEauthorblockA{\textit{Michigan Technological University}\\
Houghton, MI, USA\\
ten@mtu.edu}
\and
\IEEEauthorblockN{Yi Hu}
\IEEEauthorblockA{\textit{Michigan Technological University}\\
Houghton, MI, USA\\
yhu6@mtu.edu}
}

\begin{document}
\maketitle

\begin{abstract}
    
Accurate modeling and simulation of cooling electricity demand is essential for characterizing AI data center loads, but time-synchronized operational measurements are scarce. Constant coefficient-of-performance models can oversimplify cooling behavior, while detailed thermal state-space models are cumbersome to use. This paper proposes a closed-loop simulation framework that couples a linear thermal dynamic model with deadband-based control to capture nonlinear cooling behaviors. The framework is validated against operational data from the Marconi100 supercomputer. Compared with the baseline, the proposed model reduces the mean absolute error from 95.80 to 23.52~kW and the root-mean-square error from 109.79 to 29.91~kW. The statistical evaluation further shows improved reproduction of daily peak demand and intraday variability. The framework provides a computationally tractable approach to generate physically interpretable cooling load profiles for power system studies.
    
\end{abstract}


\section{Introduction}

The rapid expansion of artificial intelligence (AI) is driving the deployment of data centers with large and variable electricity demands. As these facilities reach capacities ranging from tens to hundreds of megawatts, characterization of their load profiles becomes important for power system planning, electricity market participation, and reliable demand response.
The total electricity demand of an AI data center comprises not only the power consumed by computing equipment but also that required by supporting infrastructure, particularly the cooling system \cite{Chen2025AIGrid}. Through coordinated control and the use of inherent thermal inertia, cooling systems can provide operational flexibility without affecting computing workloads, thereby supporting load balancing and other grid services. 
However, time-synchronized measurements of cooling system electricity consumption in AI data centers are rarely available to the public, limiting the development and validation of load characterization methods. Moreover, measurements from an individual facility reflect its specific workloads, equipment configuration, and cooling architecture, and cannot represent the diverse scenarios required for power system studies.
Therefore, explicit modeling and simulation of cooling demand are essential for characterizing AI data center loads and evaluating their impacts and flexibility in future power systems.

Cooling demand in AI data centers is often estimated by assuming a fixed relationship between cooling and computing loads, commonly represented by the coefficient of performance (COP) \cite{anka2022performance}. Although computationally convenient, this approach cannot capture the dynamic interactions among computing heat generation, ambient temperature, and cooling system operation. State-space thermal models \cite{Lu2012HVAC} provide a useful tool for residential heating, ventilation, and air conditioning (HVAC) modeling. However, they do not work for hybrid air- and liquid-cooled AI data centers whose thermal behavior is driven by both computing heat and ambient conditions. Existing data center thermal state-space models \cite{liu2022real} have been developed primarily for temperature prediction, while other studies have investigated coupled thermal energy dynamics using detailed physical models \cite{dynamiccoupling2022}. But these approaches require many system parameters that are not readily available from operational telemetry. Furthermore, although existing studies have modeled thermal behaviors, the generation of cooling electrical load profiles for power system studies remains lacking. 

Accordingly, this paper proposes a closed-loop thermal dynamic model that couples linear thermal dynamics with nonlinear deadband-based control to capture the cooling behavior and simulate the cooling electric power consumption of an AI data center. The model establishes a physically interpretable relationship between AI workload variation, ambient temperature, and cooling electric consumption.
The contributions of this paper are threefold:

\begin{itemize}[nosep,leftmargin=1.2em]
    \item \textit{First}, a reduced-order RC thermal model is developed for hybrid-cooled data centers that represents IT heat generation, ambient heat exchange, separate air and liquid cooling paths, and liquid-side heat rejection.
    \item \textit{Second}, we integrate the thermal model with binary and staged deadband control and thermal-to-electrical conversion to generate configurable cooling load profiles.
    \item \textit{Third}, we investigate different parameter estimation methods to fit the model to the CINECA Marconi100 supercomputer data \cite{Borghesi2023M100}, evaluating cooling power accuracy, daily load statistics, and internal-temperature consistency.
\end{itemize}

The source code and sample data are released to support reproducibility and benchmarking.

\label{sec:model}
\begin{figure*}[!t]
    \vspace{-6pt}
    \centering
    \includegraphics[
    width=1.0\linewidth]{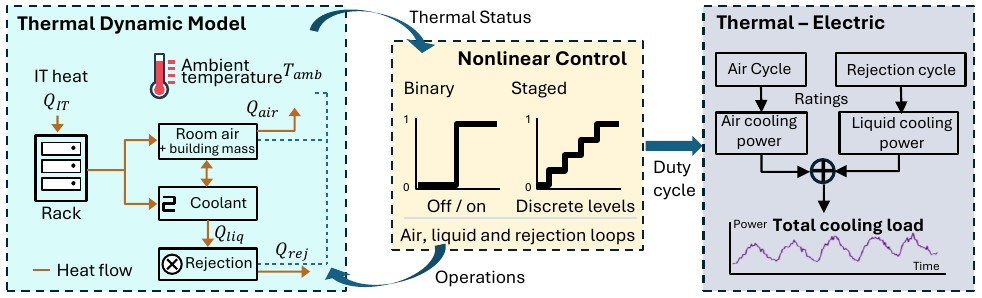}
    \caption{Overview of closed-loop framework coupling thermal dynamics, cooling control, and thermal-to-electric conversion to generate cooling load profiles.}
    \label{fig:system}
    \vspace{-6pt}
\end{figure*}

\section{Methodology}

Fig.~\ref{fig:system} illustrates the closed-loop interaction between the thermal dynamic model, nonlinear control, and thermal-electrical conversion. The parameters and control settings can be configured to represent different facility designs and operating conditions. Each of the components will be introduced in the rest of this section.

\subsection{Thermal Dynamics Models}

\begin{figure}[!b]
    \vspace{-6pt}
    \centering
    \includegraphics[width=0.8\linewidth]{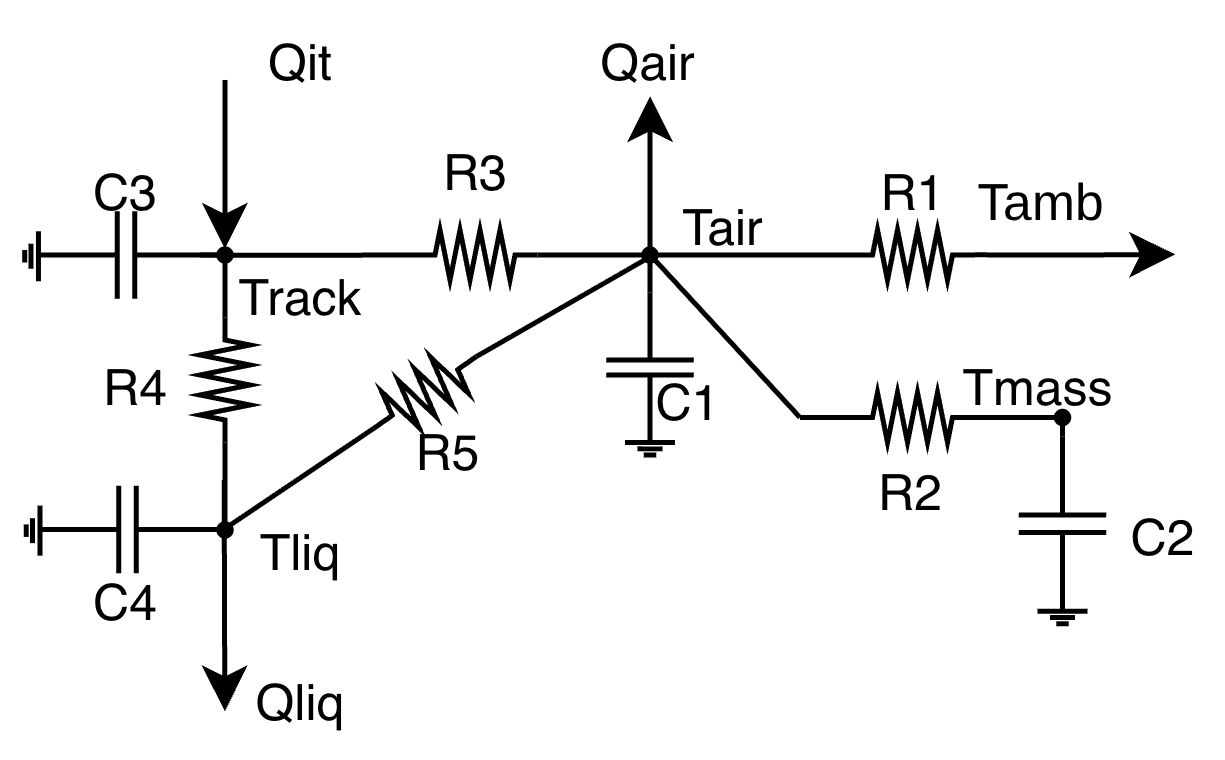}
    \caption{Four-state RC thermal dynamic model for AI data center.}
    \label{fig:proposed_model}
    \vspace{-6pt}
\end{figure}

A thermal dynamic model is extended to simulate the thermal behavior in AI data centers, as shown in Fig.~\ref{fig:proposed_model}. The heat is not only transferred from the ambient environment, but also generated by the computing components.  The state-space representation of the proposed model is given by

\vspace{-1.0em}
\begin{equation}
\begin{aligned}
    \dot{\boldsymbol{x}}=\boldsymbol{A}\boldsymbol{x}+\boldsymbol{B}\boldsymbol{u} \\
    \boldsymbol{y}=\boldsymbol{C}\boldsymbol{x}+\boldsymbol{D}\boldsymbol{u}
\end{aligned}
\label{eq1}
\end{equation}
\vspace{-0.8em}

\noindent where the thermal state of an AI data center is represented by a four-state RC network with state and input vectors

\vspace{-1.0em}
\begin{equation}
\begin{aligned}
\boldsymbol{x}=
\begin{bmatrix}
T_{\mathrm{rack}}, T_{\mathrm{air}}, T_{\mathrm{mass}}, T_{\mathrm{liq}}
\end{bmatrix}^{\!\top},\\
\boldsymbol{u}=
\begin{bmatrix}
T_{\mathrm{amb}}, Q_{\mathrm{IT}}, Q_{\mathrm{air}}, Q_{\mathrm{liq}}
\end{bmatrix}^{\!\top},
\end{aligned}
\end{equation}
\vspace{-0.5em}

\noindent where $T_{\mathrm{rack}}$, $T_{\mathrm{air}}$, 
$T_{\mathrm{mass}}$, and $T_{\mathrm{liq}}$ denote the rack, room air, thermal mass, and liquid coolant temperatures, respectively.
$T_{\mathrm{amb}}$ is the ambient temperature. The heat inputs $Q_{\mathrm{IT}}$, $Q_{\mathrm{air}}$, and $Q_{\mathrm{liq}}$ represent the Information technology (IT) heat generation, heat removed by the air cooling system, and heat removed by the liquid cooling system,
respectively. The heat flows at the four state/temperature nodes can be calculated as

\vspace{-1.5em}
\begin{equation}
\begin{aligned}
C_3\dot{T}_{\mathrm{rack}}
&=
Q_{\mathrm{IT}}
-\frac{T_{\mathrm{rack}}-T_{\mathrm{air}}}{R_3}
-\frac{T_{\mathrm{rack}}-T_{\mathrm{liq}}}{R_4},
\\
C_1\dot{T}_{\mathrm{air}}
&=
\frac{T_{\mathrm{rack}}-T_{\mathrm{air}}}{R_3}
+\frac{T_{\mathrm{amb}}-T_{\mathrm{air}}}{R_1}
\\
&\quad
+\frac{T_{\mathrm{mass}}-T_{\mathrm{air}}}{R_2}
+\frac{T_{\mathrm{liq}}-T_{\mathrm{air}}}{R_5}
-Q_{\mathrm{air}},
\\
C_2\dot{T}_{\mathrm{mass}}
&=
\frac{T_{\mathrm{air}}-T_{\mathrm{mass}}}{R_2},
\\
C_4\dot{T}_{\mathrm{liq}}
&=
\frac{T_{\mathrm{rack}}-T_{\mathrm{liq}}}{R_4}
+\frac{T_{\mathrm{air}}-T_{\mathrm{liq}}}{R_5}
-Q_{\mathrm{liq}},
\end{aligned}
\label{eq:thermal_dynamics}
\end{equation}

\noindent where the parameters $C_1$--$C_4$ denote the equivalent thermal capacitances, while $R_1$--$R_5$ denote the equivalent thermal resistances associated with the corresponding heat transfer paths. Thus, the four matrices in (\ref{eq1}) can be written as

\vspace{-1.5em}
\begin{equation}
\resizebox{0.9\columnwidth}{!}{$
\boldsymbol{A} =
\begin{bmatrix}
-\dfrac{1}{C_3}\left(\dfrac{1}{R_3}+\dfrac{1}{R_4}\right)
& \dfrac{1}{C_3R_3}
& 0
& \dfrac{1}{C_3R_4}
\\[6pt]
\dfrac{1}{C_1R_3}
& -\dfrac{1}{C_1}\left(
  \dfrac{1}{R_1}+\dfrac{1}{R_2}
  +\dfrac{1}{R_3}+\dfrac{1}{R_5}\right)
& \dfrac{1}{C_1R_2}
& \dfrac{1}{C_1R_5}
\\[6pt]
0
& \dfrac{1}{C_2R_2}
& -\dfrac{1}{C_2R_2}
& 0
\\[6pt]
\dfrac{1}{C_4R_4}
& \dfrac{1}{C_4R_5}
& 0
& -\dfrac{1}{C_4}\left(
  \dfrac{1}{R_4}+\dfrac{1}{R_5}\right)
\end{bmatrix},
$}
\end{equation}

\begin{equation}
\boldsymbol{B} =
\begin{bmatrix}
0 & \dfrac{1}{C_3} & 0 & 0
\\[6pt]
\dfrac{1}{C_1R_1} & 0 & -\dfrac{1}{C_1} & 0
\\[6pt]
0 & 0 & 0 & 0
\\[6pt]
0 & 0 & 0 & -\dfrac{1}{C_4}
\end{bmatrix},
\end{equation}

\begin{equation}
\boldsymbol{C}=\mathbf{I}_{4\times4},
\qquad
\boldsymbol{D}=\mathbf{0}_{4\times4}.
\end{equation}

\begin{figure}[!t]
    \centering
    \includegraphics[
    width=0.7\linewidth]{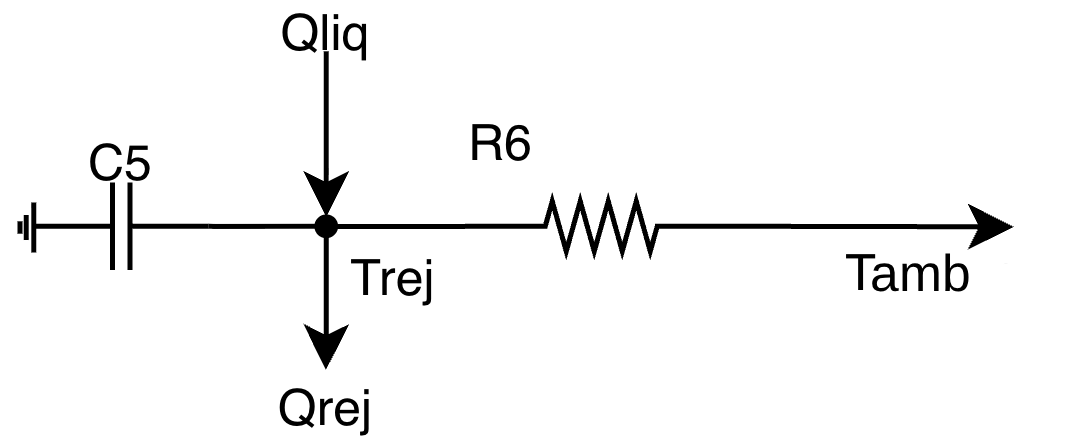}
    \caption{Single-state RC thermal dynamic model of the liquid heat rejection.}
    \label{fig:proposed_model2}
    \vspace{-8pt}
\end{figure}

\begin{algorithm}[!b]
\caption{Closed-Loop Staged Deadband Simulation}
\label{alg:cooling_staged}
\small
\begin{algorithmic}[1]

\Require Computing heat generation $Q_{\mathrm{IT}}(k)$ and ambient heat transfer $T_{\mathrm{amb}}(k)$,
\Ensure Thermal status $Q(k+1)$ and cooling command $d(k+1)$

\State Initialize $\mathbf{x}(1)$, $T_{\mathrm{rej}}(1)$,

\For{$k=1$ to $N-1$}

    \State Form $\mathbf{u}(k)$ from
    $T_{\mathrm{amb}}(k)$, $Q_{\mathrm{IT}}(k)$,
    $Q_{\mathrm{air}}(k)$, $Q_{\mathrm{liq}}(k)$

    \State $\mathbf{x}(k+1)\gets\mathbf{x}(k)
    +[\mathbf{A}\mathbf{x}(k)+\mathbf{B}\mathbf{u}(k)]\Delta t$

    \State $\displaystyle
    T_{\mathrm{rej}}(k+1)\gets T_{\mathrm{rej}}(k)
    +\dot{T}_{\mathrm{rej}}(k)\Delta t$

    \For{$\ell\in\{\mathrm{air},\mathrm{liq},\mathrm{rej}\}$
    with temperature $T_{\ell}(k+1)$}
        \State $T_{\ell}^{\mathrm{-}}\gets
        T_{\ell}^{\mathrm{set}}-\Delta T_{\ell}^{\mathrm{db}}/2$
        \State $T_{\ell}^{\mathrm{+}}\gets
        T_{\ell}^{\mathrm{set}}+\Delta T_{\ell}^{\mathrm{db}}/2$
        \If{$T_{\ell}(k+1)\leq T_{\ell}^{\mathrm{-}}$}
            \State $d_{\ell}(k+1)\gets 0$
        \ElsIf{$T_{\ell}(k+1)\geq T_{\ell}^{\mathrm{+}}$}
            \State $d_{\ell}(k+1)\gets 1$
        \Else
            \State $\displaystyle
            r\gets\frac{T_{\ell}(k+1)-T_{\ell}^{\mathrm{-}}}
            {T_{\ell}^{\mathrm{+}}-T_{\ell}^{\mathrm{-}}}$
            \State $d_{\ell}(k+1)\gets
            \lceil r\,N_{\ell}\rceil / N_{\ell}$
        \EndIf
    \EndFor

    \State $Q_{\mathrm{air}}(k+1)\gets
    Q_{\mathrm{air}}^{\mathrm{rated}}d_{\mathrm{air}}(k+1)$
    \State $Q_{\mathrm{liq}}(k+1)\gets
    Q_{\mathrm{liq}}^{\mathrm{rated}}d_{\mathrm{liq}}(k+1)$
    \State $Q_{\mathrm{rej}}(k+1)\gets
    Q_{\mathrm{rej}}^{\mathrm{rated}}d_{\mathrm{rej}}(k+1)$


\EndFor
\end{algorithmic}
\end{algorithm}

Fig.~\ref{fig:proposed_model2}
represents the liquid heat rejection process where the indoor heat removed by the liquid coolant, denoted by $Q_{\mathrm{liq}}$, enters the heat rejection subsystem. A portion of this heat may be transferred to the ambient environment through the equivalent thermal resistance $R_6$, while the remaining heat is rejected by the cooling equipment. 
The energy balance of the heat rejection subsystem is expressed as

\begin{equation}
C_{\mathrm{5}}
\dot{T}_{\mathrm{rej}}
=
Q_{\mathrm{liq}}
+
\frac{T_{\mathrm{amb}}-T_{\mathrm{rej}}}
     {R_{\mathrm{6}}}
-
Q_{\mathrm{rej}},
\label{eq:rejection}
\end{equation}

\noindent where $T_{\mathrm{rej}}$ is the effective rejection side temperature,
$C_{\mathrm{5}}$ is the associated thermal capacitance.



\subsection{Deadband-Based Nonlinear Control}
\label{sec:control_calibration}

Building on the thermal dynamic model, Algorithm~\ref{alg:cooling_staged} couples temperature evolution with cooling control to generate thermal heat-removal rates and the duty cycles of air-liquid cooling facilities. Each controller $\ell\in\{\mathrm{air},\mathrm{liq},\mathrm{rej}\}$ 
determines a normalized capacity command $d_{\ell}(k)\in[0,1]$ using either binary or staged control.

\paragraph{Binary on/off control.}
The binary controller retains its previous operating state between the band limits:
\begin{equation}
d_{\ell}(k)=
\begin{cases}
1, & T_{\ell}(k)\ge T_{\ell}^{+},\\
0, & T_{\ell}(k)\le T_{\ell}^{-},\\
d_{\ell}(k-1), & \text{otherwise},
\end{cases}
\label{eq:onoff}
\end{equation}

\noindent where $T_{\ell}^{\mathrm{set}}$ is the set temperature and $\Delta T_{\ell}^{\mathrm{db}}>0$ is the dead-band width, with limits
\begin{equation}
T_{\ell}^{-}
=T_{\ell}^{\mathrm{set}}-\tfrac12\Delta T_{\ell}^{\mathrm{db}},
\qquad
T_{\ell}^{+}
=T_{\ell}^{\mathrm{set}}+\tfrac12\Delta T_{\ell}^{\mathrm{db}}.
\label{eq:band}
\end{equation}
The equipment operates at either zero or rated capacity.

\paragraph{Staged control.}
For $N_{\ell}\ge2$ equally spaced nonzero stages, the controller maps temperature to a discrete capacity fraction:
\begin{equation}
d_{\ell}(k)=
\begin{cases}
0, & T_{\ell}(k)\le T_{\ell}^{-},\\[3pt]
\displaystyle
\frac{1}{N_{\ell}}
\left\lceil
N_{\ell}
\frac{T_{\ell}(k)-T_{\ell}^{-}}
     {T_{\ell}^{+}-T_{\ell}^{-}}
\right\rceil,
& T_{\ell}^{-}<T_{\ell}(k)<T_{\ell}^{+},\\[9pt]
1, & T_{\ell}(k)\ge T_{\ell}^{+}.
\end{cases}
\label{eq:staged}
\end{equation}
Thus, $d_{\ell}(k)\in\{0,1/N_{\ell},\ldots,1\}$.
For example, four stages provide capacity fractions of $25\%$, $50\%$, $75\%$, and $100\%$, in addition to the off state.

\subsection{Thermal and Electrical Energy Conversion}
The capacity commands determine both the thermal status of the next step and the electric power consumed level in this step. The air-side, liquid-side, and total cooling electrical demands are
\begin{equation}
\begin{aligned}
P_{\mathrm{air}}(k)
&=d_{\mathrm{air}}(k)P_{\mathrm{air}}^{\mathrm{rated}},\\
P_{\mathrm{liq}}(k)
&=d_{\mathrm{rej}}(k)P_{\mathrm{liq}}^{\mathrm{rated}},\\
P_{\mathrm{cool}}(k)
&=P_{\mathrm{air}}(k)+P_{\mathrm{liq}}(k).
\end{aligned}
\label{eq:duty_P}
\end{equation}

\noindent Their proportional scaling assumes a fixed ratio of thermal capacity to electrical demand for each paired output while operating. The temperature-dependent switching or quantization makes the overall closed-loop model nonlinear, although the thermal state equations and output scaling are linear.

\subsection{Parameter Identification}
\label{sec:param_identification}

This section provides guidelines to estimate parameters for the proposed simulation framework.

\subsubsection{Physics-Informed Parameter Initialization}

The proposed model parameters are adapted from parameters of the residential HVAC model in~\cite{Lu2012HVAC}. The residential model provides the initial physical relationships for the air and thermal-mass capacitances and envelope resistances:
\begin{equation}
\begin{aligned}
C_1 &= V\rho c_p,
\qquad V=Ah,
\qquad C_2=m_{\mathrm{mass}}c_p,
\\
R_1 &= \frac{1}{\mathrm{UA}},
\qquad
R_2 = \frac{1}{\mathrm{UA}_{\mathrm{mass}}},
\end{aligned}
\label{eq:physics_init}
\end{equation}

\noindent where \(A=2.5\times10^4\,\mathrm{m}^2\), \(\mathrm{UA}=5391\,\mathrm{W/K}\), and \(\mathrm{UA}_{\mathrm{mass}}=4.3\times10^5\,\mathrm{W/K}\). The air capacitance is recomputed using \(c_p=1005\,\mathrm{J/(kg\cdot K)}\). The proposed model further extends the residential structure with a rack node and a liquid-cooling node, introducing additional thermal capacitances and heat-transfer resistances.


Parameters that are not available from the residential model or M100 dataset \cite{Borghesi2023M100} are initialized using engineering assumptions and estimations. 
Rack-path resistances are estimated from steady-state temperature differences over the first seven days of the selected telemetry segment:
\begin{equation}
\begin{aligned}
R_3 &\approx
\mathrm{median}\!\left(
\frac{T_{\mathrm{rack}}-T_{\mathrm{air}}}
{\max(Q_{\mathrm{IT}}-Q_{\mathrm{liq}},0)}
\right),
\\
R_4 &\approx
\mathrm{median}\!\left(
\frac{T_{\mathrm{rack}}-T_{\mathrm{liq,mid}}}
{Q_{\mathrm{liq}}}
\right),
\end{aligned}
\label{eq:R34}
\end{equation}
where \(T_{\mathrm{liq,mid}}=\tfrac12(T_{\mathrm{liq,supply}}+T_{\mathrm{liq,return}})\). The weak liquid-to-air path \(R_5\) is assigned a large initial value to limit direct heat transfer between the two cooling paths.
Temperature setpoints are initialized from median measured temperatures, and deadband widths from the inter-percentile range \(p_{90}-p_{10}\). 

The remaining capacitances, resistances, deadband widths, staged-control settings, and rated powers are initialized within physically plausible ranges. Calibration is performed using Algorithm~\ref{alg:cooling_staged} over winter and summer calibration windows (2022-03-17 and 2022-07-17). The resulting physics-informed parameter set is denoted by \(\theta_0\).

\subsubsection{Maximum-Likelihood Estimation (MLE)}

Following the physics-informed initialization, a subset of the parameters is further refined using MLE:

\vspace{-1.0em}
\begin{equation}
\theta =
\bigl[
R_1,R_2,R_3,R_4,C_1,C_2,C_3,C_4
\bigr].
\label{eq:theta_free}
\end{equation}

\noindent The remaining parameters are held at their initialized values. 
The observed output \(y_k\) is modeled as the simulated output plus independent and identically distributed Gaussian measurement noise:

\vspace{-1.0em}
\begin{equation}
z_k =
\hat{z}_k(\theta)+\varepsilon_k,
\qquad
\varepsilon_k\sim\mathcal{N}(0,\sigma^2).
\end{equation}

\noindent The objective of maximizing the likelihood is equivalent to minimizing the sum of squared errors (SSE). We have explored two objective functions:

\paragraph{Objective A: Total cooling power}
The first objective minimizes the error between measured and simulated total cooling electric power
\begin{equation}
J(\theta)
=
\sum_k
\left(
P_{\mathrm{cool},k}^{\mathrm{meas}}
-
P_{\mathrm{cool},k}^{\mathrm{sim}}(\theta)
\right)^2.
\label{eq:j_pcool}
\end{equation}

\paragraph{Objective B: Air--liquid power split}
The second objective considers the air and liquid cooling electrical power separately. Let \(s_{\mathrm{air}}\) and \(s_{\mathrm{liq}}\) denote the standard deviations of the corresponding measured load. The weighted objective is
\begin{equation}
J_\lambda(\theta)
=
\lambda
\frac{\mathrm{SSE}_{\mathrm{air}}(\theta)}
{s_{\mathrm{air}}^2}
+
(1-\lambda)
\frac{\mathrm{SSE}_{\mathrm{liq}}(\theta)}
{s_{\mathrm{liq}}^2},
\label{eq:j_lambda}
\end{equation}
where \(\lambda\in\{0.1,0.2,0.3\}\) controls the weights of the two cooling paths.

The deadband-based and staged switching logic makes the mapping from parameters to objective value, \(\theta\mapsto J(\theta)\), nonsmooth. Therefore, differential evolution (DE) is used instead of gradient-based nonlinear least squares. Each parameter is constrained by bounds of \([0.1,10]\times\theta_0\). 



\section{Simulation Results}
\label{sec:results}

\subsection{Simulation Setup and Baseline Selection}

The proposed model is validated using operational telemetry from the Marconi100 supercomputer \cite{Borghesi2023M100}. Throughout this section, ``\textit{measured}'' denotes the actual cooling power from the M100 dataset. The simulation is performed at a 1-s resolution and then downsampled to 15-min intervals to match the resolution of the measured data.

The baseline is the static mapping from IT heat to cooling load implied by a COP, the ratio of heat removed to the electrical work required to remove it, denoted as $P_{\mathrm{cool,baseline}}= \frac{Q_{\mathrm{IT}}}{\mathrm{COP}}$.
Temperature-dependent COP curves are often tied to Computer Room Air Conditioner (CRAC) supply air temperature and specific air-cooled configurations, which limits their transferability to hybrid cooling systems \cite{bai2018comparative}. More detailed dynamic models explicitly represent individual cooling components and their operating parameters \cite{dynamiccoupling2022,Wang2026MOMPC}. However, these models rely on system-specific cooling configurations that are not directly available to the system considered here, and they do not generate cooling load profiles. Thus, the constant-COP model is selected as the baseline. 


\subsection{One-on-One Evaluation on Cooling Load Profile}
\begin{figure}
    \centering
    \includegraphics[width=\linewidth]{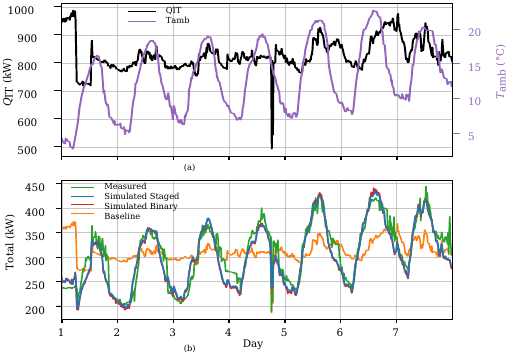}
    \caption{Example of simulation result: (a) Inputs: QIT and Tamb (b) measured, simulated, and baseline weekly cooling load profiles.}
    \label{fig:one_on_one_results}
    \vspace{-6pt}
\end{figure}

The example of the measured, proposed model simulated (with two control methods), and the baseline cooling load profiles are shown in Fig.~\ref{fig:one_on_one_results} (b) over one week at 15-minute granularity, with the corresponding computing load and ambient temperature curves shown in Fig.~\ref{fig:one_on_one_results} (a). By visual inspection, the simulated profile follows the measured load much better than the baseline model.



To quantitatively compare the simulated cooling load with the measured data and the constant COP baseline, the mean absolute error (MAE) and the root mean square error (RMSE) are used to measure the simulation accuracy, and Pearson's correlations $\mathrm{Corr}(Q_{\mathrm{IT}},P_{\mathrm{cool}})$ are used to characterize the dependence on IT load. The results in table~\ref{tab:results} shows that the proposed model substantially reduces both errors and reproduces the measured correlation more closely, demonstrating its ability to capture thermal dynamics not represented by the baseline.

\begin{table}
\caption{Quantitative Evaluation of Simulated Cooling Load}
\label{tab:results}
\centering

\begin{tabular}{lccc}
\toprule
Model & MAE (kW) & RMSE (kW) & $\mathrm{Corr}(Q_{\mathrm{IT}},P_{\mathrm{cool}})$ \\
\midrule
Measured & - & - & $0.224$ \\
Baseline & 95.80 & 109.79 & $1.0$ \\
Simulated (binary) & 24.21 & 30.27 & $0.280$ \\
Simulated (staged) & 23.52 & 29.91 & $0.282$ \\
\bottomrule
\end{tabular}%
\vspace{-8pt}
\end{table}



\subsection{Statistical Evaluation on Simulated Dataset}


\begin{figure}
    \centering
    \includegraphics[width=\linewidth]{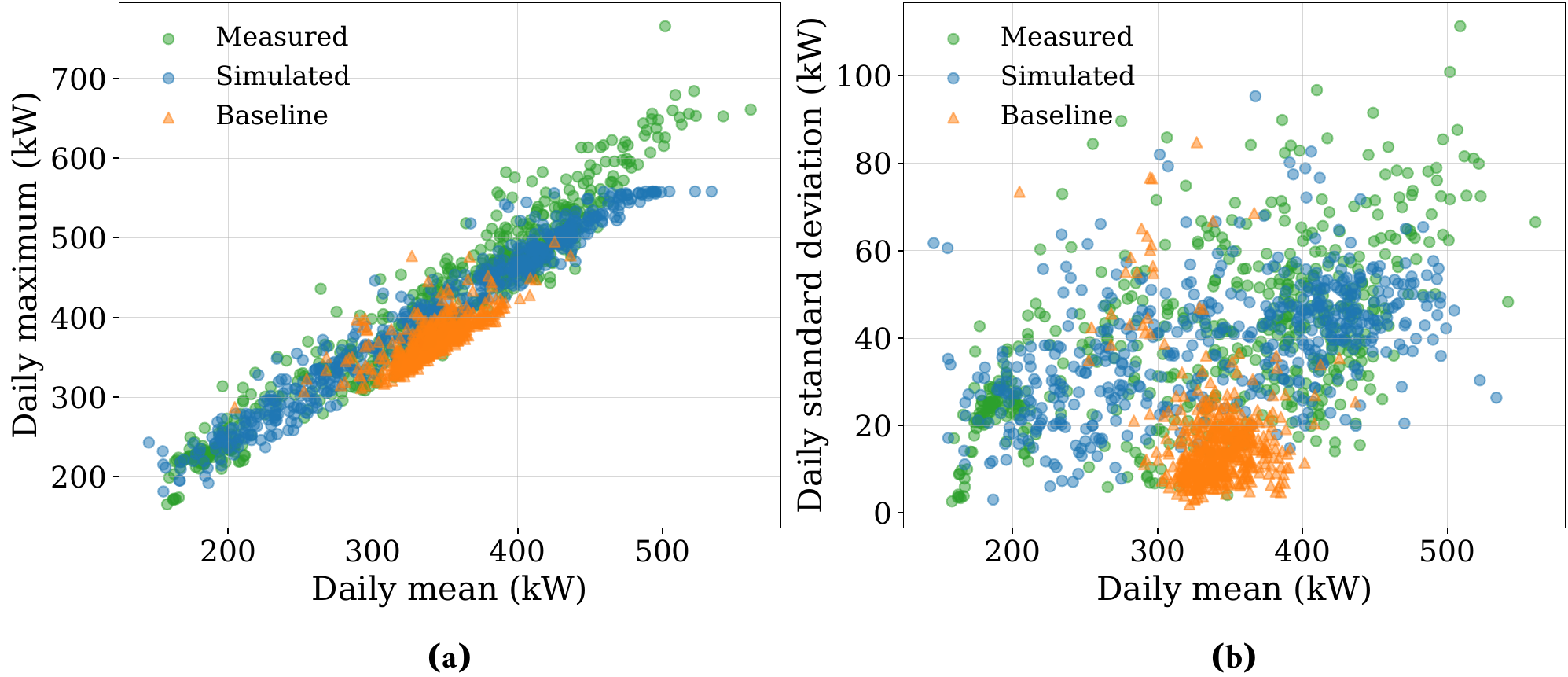}
    \caption{Statistical comparison over 520 days: (a) daily mean vs. daily peak; (b) daily mean vs. daily std.}
    \label{fig:stat_results}
    \vspace{-6pt}
\end{figure}

To evaluate statistical fidelity across diverse operating conditions, 520 daily cooling load profiles are compared using the daily mean, maximum, and standard deviation, with each point in Fig.~\ref{fig:stat_results}(a) and (b) representing one day. In the mean–maximum comparison, the simulated profiles follow the measured trend across most operating levels, whereas the baseline exhibits a narrower distribution and underrepresents the variation in daily peaks. In the mean–standard-deviation comparison, the proposed model captures substantially more of the measured intraday variability than the baseline. These results demonstrate that the proposed model better reproduces both daily peak behavior and temporal variability.


\subsection{Evaluation on Different Parameter Estimation Methods}
\label{sec:mle_results}

Table~\ref{tab:mle_metrics} shows that the choice of parameter estimation method affects the performance:

\begin{itemize}[nosep, leftmargin=1.2em]
    \item The physics-informed parameter estimation \(seed \theta_0\) provides a reasonable representation of total cooling power and the highest accuracy for rack temperature simulation, providing a reference for evaluating the effect of parameter optimization on the internal thermal states.
    \item The total-power MLE objective seeks to reduce the error in \(P_{\mathrm{cool}}\). However, compared with the physics-informed initialization, this fit yields higher reported errors for total, air, and liquid cooling power, as well as rack temperature. Furthermore, the estimated \(R_3\) increases to approximately \(9.0\) times its initial value, approaching the upper optimization bound. These results suggest that fitting total cooling power alone does not adequately constrain individual cooling paths or internal thermal dynamics
    \item Additional MLE experiments split the air and liquid cooling load using weighting factors \(\lambda\in\{0.1,0.2,0.3\}\). Among the tested configurations, \(\lambda=0.2\) provides the best balance between the air and liquid power predictions. However, this configuration produces a larger rack temperature error. This indicates that this configuration can improve the electrical power fit without preserving the accuracy of the internal thermal states.
\end{itemize}

Overall, the results demonstrate that different objective functions produce different parameter estimates and trade-offs between electrical power accuracy and thermal-state consistency. The physics-informed seed retains the best rack temperature accuracy, while the MLE configurations improve selected electrical power predictions. This highlights the importance of considering both electrical and thermal performance when identifying parameters for the proposed framework.

\begin{table}[!t]
\centering
\caption{Results with Different Parameter Estimation Methods}
\label{tab:mle_metrics}
\scriptsize
\setlength{\tabcolsep}{3pt}

\begin{tabular*}{\columnwidth}{@{\extracolsep{\fill}}llcc@{}}
\toprule
Parameter Estimation Methods & Quantity & MAE & RMSE \\
\midrule
Physics-informed (Seed $\theta_0$)
  & $P_{\mathrm{cool}}$ (kW) & 23.52 & 29.91 \\
  & $P_{\mathrm{air}}$ (kW) & 12.56 & 22.63 \\
  & $P_{\mathrm{liq}}$ (kW) & 30.93 & 38.59 \\
  & Track ($^\circ$C) & 2.03 & 2.62 \\
\midrule
MLE ($P_{\mathrm{cool}}$-only)
  & $P_{\mathrm{cool}}$ (kW) & 25.18 & 32.01 \\
  & $P_{\mathrm{air}}$ (kW) & 13.55 & 23.54 \\
  & $P_{\mathrm{liq}}$ (kW) & 33.41 & 41.14 \\
  & Track ($^\circ$C) & 15.47 & 15.74 \\
\midrule
MLE (Power split $\lambda=0.2$)
  & $P_{\mathrm{cool}}$ (kW) & 22.59 & 28.53 \\
  & $P_{\mathrm{air}}$ (kW) & 13.42 & 22.99 \\
  & $P_{\mathrm{liq}}$ (kW) & 30.54 & 38.21 \\
  & Track ($^\circ$C) & 4.19 & 4.75 \\
\bottomrule
\end{tabular*}
\vspace{-6pt}
\end{table}

\section{Conclusion}

This paper presented a closed-loop framework for simulating AI data center cooling loads. Validation against Marconi100 operational data reduced the mean absolute error from 95.80 to 23.52 kW and the root-mean-square error from 109.79 to 29.91 kW relative to the baseline. Evaluation across 520 daily profiles also demonstrated improved reproduction of daily peak demand and intraday variability. Different parameter estimation methods trade-offs between electric load accuracy and thermal-state consistency. Future work will focus on validation across additional cooling configurations, and integration with computing workload models to support studies of data center flexibility and grid interaction.

\bibliographystyle{IEEEtran}
\bibliography{reference}

\end{document}